\def\year{2017}\relax
\documentclass[letterpaper]{article}
\pdfoutput=1
\usepackage{aaai17}

\usepackage{helvet}  
\usepackage{courier} 
\usepackage{xcolor}
\usepackage{soul}
\usepackage{subfig}
\usepackage{graphicx}
\usepackage{multirow}
\usepackage{url}
\usepackage{arydshln} 
\usepackage{balance}
\usepackage{times}
\usepackage{setspace}
\usepackage{comment}
\usepackage{enumerate}
\usepackage{textcomp}
\usepackage{booktabs}
\usepackage{ccicons}
\usepackage{todonotes}

\newcommand{\me}{\sf \small About.me}
\newcommand{\lnkd}{\sf \small LinkedIn}
\newcommand{\fb}{\sf \small Facebook}
\newcommand{\ing}{\sf \small Instagram}
\newcommand{\twt}{\sf \small Twitter}

\newcommand{\flk}{\sf \small Flickr}
\usepackage{ntheorem}
\usepackage{amssymb,amsmath}
\usepackage{latexsym}   
{
\theoremheaderfont{\sc}
\theorembodyfont{\slshape}
\theorembodyfont{\upshape}
\theoremseparator{}
\theoremnumbering{arabic}
\theoremsymbol{\ensuremath{\Box}}
\newtheorem{myPro}{Problem}
}

\newcommand\fref[1]{Figure~\ref{#1}}
\newcommand\tref[1]{Table~\ref{#1}}

\newcommand\etal{~\textsl{et al.}} 
\newcommand\tabhead[1]{\small\textbf{#1}}

\newcommand{\subtitle}[1]{ \noindent \textbf{#1} } 

\makeatletter

\begin{document}
\setcounter{secnumdepth}{0}
\title{Wearing Many (Social) Hats:\\How Different are Your Different Social Network Personae?\thanks{This work was partially supported by the Space for Sharing (S4S) project (Grant No. ES/M00354X/1). This research was also in part supported by NSF CNS-1422215, NSF IUSE-1525601, and Samsung GRO 2015 awards.}
\\ {\Large \normalfont [Please cite the ICWSM'17 version of this paper]}
}

\frenchspacing
\setlength{\pdfpagewidth}{8.5in}
\setlength{\pdfpageheight}{11in}
\pdfinfo{
/Title (Wearing Many (Social) Hats: How Different are Your Different Social Network Personae?)
/Author (Changtao Zhong, Hau-wen Chan, Dmytro Karamshu, Dongwon Lee, Nishanth Sastry)}

\author{
		Changtao Zhong\thanks{Work done while at King's College London}{\small $^1$}, Hau-wen Chang\thanks{Work done while at Penn State University}{\small $^2$} , Dmytro Karamshuk\thanks{Work done while at King's College London}{\small $^3$}, Dongwon Lee{\small $^4$}, Nishanth Sastry{\small $^5$}\vspace{0.05in}\\
$^1$ Twitter, UK 
$^2$ IBM Watson, US 
$^3$ Skyscanner, UK \\ 
$^4$ College of Information Sciences and Technology, The Pennsylvania State University, US\\
$^5$ Department of Infromatics, King's College London, UK 
\vspace{0.05in}\\
$^1$ czhong@twitter.com 
$^2$ changha@us.ibm.com
$^3$ dima.karamshuk@skyscanner.net\\
$^4$ dlee@ist.psu.edu
$^5$ nishanth.sastry@kcl.ac.uk
}

\maketitle
\begin{abstract}
This paper investigates when users create profiles in different social networks, whether they are redundant expressions of the same persona, or they are adapted to each platform. Using the personal webpages of 116,998 users on {\me}, we identify and extract matched user profiles on several major social networks including {\fb}, {\twt}, {\lnkd}, and {\ing}. We find evidence for distinct site-specific norms, such as differences in the language used in the text of the profile self-description, and the kind of picture used as profile image. By learning a model that robustly identifies the platform given a user's profile image (0.657--0.829 AUC) or self-description (0.608--0.847  AUC), we confirm that users do adapt their behaviour to individual platforms in an identifiable and learnable manner. However, different genders and age groups adapt their behaviour differently from each other, and these differences are, in general, consistent across different platforms. We show that differences in social profile construction correspond to differences in how formal or informal the platform is.
\end{abstract}

\section{Introduction}
\label{sec-1}
As social network usage becomes more common, new platforms with large user bases have multiplied, and users are increasingly creating multiple profiles across different platforms. 
The latest survey from the Pew Research Center reported a sharp rise in multi-platform usage: more than half of all online adults  use two or more social networks~\cite{duggan2016social}. 
However, social networks have traditionally been \emph{walled gardens}, and information within a site is not usually exported to other sites. 
Consequently, using multiple social networks entails the creation of  different social network profiles on each platform, and has led to practices such as cross posting of the same information across networks~\cite{lim2015mytweet}. 

In this paper, we are interested in the  fundamental act of creating a social network profile on different platforms. Profiles can be seen as a form of ``implicit'' identity construction~\cite{zhao2008identity}, and our primary motivation is to see how this identity is negotiated or adapted across different platforms. We attempt to understand profile construction under the framework of Tajfel's Social Identity Theory~\cite{hogg2006social}, which suggests that the process of identity creation involves, among other things, \emph{self-categorisation} -- defining oneself in relation to other (real or perceived) groups of people, e.g., by following in-group customs, and the consequent \emph{depersonalisation} -- the transformation from being an idiosyncratic individual to being a member of a group. 

We first attempt to understand self-categorisation in terms of following \emph{platform-specific norms}. Although  major social platforms such as {\fb}, {\twt}, or {\lnkd} offer roughly similar social functionalities, there are some feature differences, and more importantly, differences in culture or expected etiquette of acceptable user behaviour~\cite{mclaughlin2012norm,walther2008role,cimino2009netiquette}. Since some behaviours are considered as inappropriate in particular contexts, perceived norms impose constraints on profile construction~\cite{mclaughlin2012norm,zhao2016social}. For instance, Facebook attempts to institute a ``real name policy''\footnote{\url{https://www.facebook.com/help/112146705538576}}, and some classes of professionals on LinkedIn may face an implicit pressure to wear suits for their profile pictures. An important question, therefore, is whether users follow platform norms and whether this leads to  differences in profile construction across platforms. 

Although there are platform-wide policies, norms, trends, and etiquettes, profile generation is a deeply personal ``explicit act of writing oneself into being in a digital environment  and participants must determine how they want to present themselves to those who may view their self-representation or those who they wish might.''~\cite{boyd:2011}. Social network profiles have been seen as ``taste performances'' -- carefully constructed expressions of user taste and personality~\cite{liu2007social}. Thus, it is interesting to investigate whether  the ways in which profiles are adapted to suit individual platforms are automatically identifiable or whether they are uniquely idiosyncratic. At the same time, \emph{group associations} have an important role to play in identity construction~\cite{mccall1978identities}, and  different audience segments have different expectations about one's public profiles~\cite{rui2013strategic}. Thus, we are interested in understanding how users from different segments (e.g., from different demographics) would express themselves differently inside the same social platform.
In summary, we organise our study of identity and profile construction across social networks as follows:
\begin{enumerate}[{\bf RQ-1 }]
	\item Do users construct their profile identities differently on different platforms? 
	\item Are cross-platform differences in profile construction consistent and identifiable?
	\item Do different social groups and demographics have different ways of presenting themselves? Are  group-specific aspects of constructing profiles  common across platforms?
\end{enumerate}

To answer these questions, and understand the relationship(s) between the different social network personae of a user, we  collected a seed set of different profiles of users on up to five  social network platforms, by making use of the personal web hosting site {\me}\footnote{The datasets used in this paper are shared for wider community use at \url{http://nms.kcl.ac.uk/netsys/datasets/aboutme}.}. On {\me}, users can create personal profiles and provide links to other social platforms. Using these links, we created a matched set of identities on different social networks, all belonging to the {\em same} user. We focus on the top-4 social networks represented in the dataset -- {\fb}, {\lnkd}, {\twt}, and {\ing} -- as representatives of today's social web.
Goffman~\cite{goffman1959presentation} suggested that the presentation of self involves both the explicit communication -- the ``given'' aspects, as well as implicit, e.g., non-verbal, communication -- the ``given-off'' aspects. To capture both, we draw on both the major aspects of user profiles, which are often used and heavily customised: (1) given: the textual self-description or self-summary of the user on their profile page, and (2) given-off: the profile image itself. 
Since both the profile image and text  can be free form, and individual users can customise almost any aspect, we focus on aspects that are generalisable across social networks. 

To the best of our knowledge, this is the first study looking at how user profiles vary across different social networks, both in the aggregate (as collections of user profiles across different networks), as well as in the individual (as a suite of profiles created by the same user across different social platforms). 
Although this is intended as a first-cut study, we believe this line of research would be crucial to understand, adapt and support the recent rapid rise of multi-platform online social network usage~\cite{duggan2016social}.

\section{Preliminaries} 
\label{sec:preliminaries}
\subsection{Dataset Construction}
\label{sec:dataset}
{\me} is a simple social media platform that provides a social directory service. Each user has an individual profile page that includes a short biography and a set of attributes including location, education, achievements, skills, and tags.  Importantly, each user can list links to their profiles in other social network sites. As users themselves voluntarily provide such links to their  ``other'' identities, the quality of such links is high and thus ideal for our intended purpose of comparing user profiles across social network sites.

We sampled seed users as follows: First, we searched for random {\me} users located in U.S. top-50 cities\footnote{While the precise location distribution of {\me} users is unknown, searches for users from non-US cities yielded scarce results.}
and collected the tags used by these seed users. 
We then extended our seed users by adding users associated with at least one of the top-200 most used tags. We crawled  {\me} profiles of all users that we detected and extracted links to their profiles on other social networks. In total, we have collected 170,348 unique {\me} users, of which 116,998 have at least one other social network account listed. 

Analyzing the distribution of links to other social networks that {\me} users listed on their profiles, we found that 76\% of {\me} users in our dataset have linked their profiles to their alternate account in {\twt}, whereas users with links to their {\lnkd}, {\fb} and {\ing} profiles took a share of 65\%, 46\% and 32\%, respectively. 
Based on this, in the following experiments, we focused our analysis around these top-4 social networks. 
From 116,998 {\me} users, next, we have crawled corresponding profiles linked to the above mentioned social networks 
and extracted the \emph{profile images} and \emph{self-descriptions} of the profiles on each site\footnote{
Since {\fb} self-description is not publicly available in its new launched API 2.0,  we excluded {\fb} in our analysis for self-description. 
}.
\tref{tab:profile-across-stats} shows the distribution of the number of profiles with corresponding images and self-descriptions successfully extracted.

\begin{table}[tb]
{\small
\center
\begin{tabular}{c|c|c|c}
\hline
\tabhead{\#} & \tabhead{Profile image} & \tabhead{\#} & \tabhead{Self-description}\\
\hline
0	&15,933 (13.6\%) &0&2,743	 (2.3\%) \\
1	&55,447 (47.4\%)&1&30,591 (26.1\%) \\
2	&36,548 (31.2\%)&2&55,598 (47.5\%)\\
3	&3,576 (3.1\%)&3&24,602 (21\%) \\
4	&2,167 (1.9\%)&4&3,457	 (3\%)\\
5    & 3,327 (2.8\%)&5&  N/A \\
\hline
        Total & 116,998  & Total & 116,998\\
\hline
\end{tabular}
\caption{
	\textbf{Distribution of the number of profiles.}  
Numbers in brackets indicate the fraction of those users among total number of {\me} users.
}
\label{tab:profile-across-stats}
}
\vspace{-0.5cm}
\end{table}

\subsection{Feature Extraction}
\label{sec:prelim-feature-extract}

\subtitle{Profile images: facial features.}
The state-of-the-art  face detection techniques from Computer Vision~\cite{viola2004robust,wright2009robust}  can achieve a very high accuracy on various datasets~\cite{jenkins2008100,jain2010fddb}. Therefore, we adopted two publicly available face detection software to analyse profile images in our dataset: Face++\footnote{\url{http://www.faceplusplus.com/detection_detect/}}~\cite{zhou2013extensive} and Microsoft Face APIs\footnote{\url{https://www.microsoft.com/cognitive-services}}. Each library takes an image as an input and produces a number of facial features (e.g., detected age, smiling score) which characterize the face (or faces) in the image.  We summarize these features in \tref{tab:api-comparison}.

\begin{table}
{\small
\centering
\begin{tabular}{c|c|c}
	\hline 
	&\tabhead{Face++} & \tabhead{Microsoft} \\ 
	\hline 
	\textbf{Faces numbers} & $\surd$ & $\surd$ \\ 
	\textbf{Face gender} & $\surd$  & $\surd$ \\
	\textbf{Face age} & $\surd$ & $\surd$ \\
	Face race &  $\surd$ &  \\
	\textbf{Face smiling} & $\surd$ & \\
	\textbf{Face glasses} & $\surd$ & \\
	Face pose (3) & $\surd$ & $\surd$ \\
	\textbf{Image category} & & $\surd$\\
	Image color & & $\surd$ \\
	\hline 
\end{tabular}
\caption{\textbf{A total of 11 features, including 3 components of face pose, were provided by Face++ and Microsoft APIs.}}
\label{tab:api-comparison}
\vspace{-0.5cm}
}
\end{table}

Previous studies on Face++~\cite{jain2010fddb,bakhshi2014faces} have reported a high accuracy for face detection (with $97\% \pm 0.75\%$ overall accuracy) using various manually labeled or crowd-sourced datasets. Our experiments on {\me} dataset also suggest a comparable performance from the Microsoft API: we performed a sanity check by testing whether the two APIs are consistent with each other using three facial features common for both APIs -- face number, gender and age. 
Our analyses on face number show that the results from two APIs were consistent for over $90\%$ of images. We next used images with a single face detected and compared the gender and the age of faces as detected by two APIs. Our results suggest that over $88.5\%$ of images are considered to be of the same gender in both APIs, and about $80\%$ 
of them have less than 10 years difference in the predicted ages. In summary, both APIs are highly consistent with each other.

\subtitle{Profile image: Deep learning Features.}
We use deep convolution networks~\cite{krizhevsky_imagenet_2012}--the state-of-the-art approach in object recognition~\cite{deng_imagenet_2009}--to extract a set of more fine-grained visual features for our machine learning tasks.  More specifically, we train a deep convolution network using Caffe library~\cite{jia2014caffe} based on 1.3 million images annotated with 1,000 ImageNet classes and apply it to classify profile images, and extract two types of visual features: (1) \emph{Deep neural network} features (with $4,096$ dimensions) from the layer right before the final classification layer, which are known for a good performance in semantic clustering of images~\cite{donahue_decaf_2013}; and (2) \emph{Recognised objects} among the 1,000 Image Object classes that the model is trained on.

\subtitle{Self-description: Word2Vec Features.}
We analyze the self-descriptions of the users with the Word2Vec methodology proposed in \cite{mikolov2013distributed,mikolov2013efficient}. More specifically, we extract the word vector representation from the corpus consisting of all collected self-descriptions in our dataset using the Gensim natural language processing library\footnote{\url{https://radimrehurek.com/gensim/}}.  We further reduce the dimensionality of the word space for our analysis as follows. Firstly, we cluster words in Word2Vec dictionary in $N = 100$ groups\footnote{The value of $N$ has been chosen by optimizing for the classification performance in the machine learning tasks from the later sections.} using k-means clustering algorithm. Each cluster is presumed to represent a group of words located nearby in the Word2Vec space and, therefore, representing different semantics. We then map each self-description in a $100$-dimensional feature vector in which each dimension represents the normalized frequency of words from each Word2Vec cluster.

\begin{table}[tb]
{\small
\centering
\subfloat[Gender]{
\centering
\begin{tabular}{cc|c}
\hline
 \tabhead{Male (M)} & \tabhead{Female (F)} & \tabhead{Total} \\
\hline
        63,012 & 41,197 & 104,209\\
\hline
\end{tabular}
\label{tab:demographic-dis-gender}
}
\\
\subfloat[Age]{
	\begin{tabular}{ccc|c}
\hline
\tabhead{Youth (Y) } & \tabhead{Young Adults (YA)} & \tabhead{Adults (A)} & \multirow{2}{*}{\tabhead{Total}} \\
\tabhead{$\le 25$ } & \tabhead{$26-34$ } & \tabhead{$\ge 35$} & \\
\hline
       11,312 & 21,059 & 29,705 & 62,076 \\
\hline
\end{tabular}
\label{tab:demographic-dis-age}
}
\caption{\textbf{The distribution of user demographics.}
}
\label{tab:demographic-dis}
}
\vspace{-0.5cm}
\end{table}

\subsection{Demographic Inference}
\label{sec:demographic}

Next, we introduce the method we used to infer the demographic information of users (see \tref{tab:demographic-dis}).

\subtitle{Gender.} To extract reliable gender information, we check the gender information detected from users' profiles images and remove images that are divergent for Face++ and Microsoft APIs.This results in 63,012 male and 41,107 female users. The higher number of males is in line with Alexa traffic statistics for the {\me} website\footnote{\url{http://www.alexa.com/siteinfo/about.me}}. We also validate the performance of the method using  data collected from {\flk}, which has gender information of 5,627 users. We find that the accuracy of our method is over 98\% for males and over 96\% for  females. 

\subtitle{Age.} In our age inference, we divide users into three groups, $\le 25$, $26-34$ and $\ge 35$, to represent Youth, Young Adults, and Adults, respectively, following the age demographic method used in market analysis~\cite{mir2012}. To decide which group a user belongs to, we calculate the average age detected from the user's all profile images, and label the user based on the average age. Removing users that are assigned different age groups by Face++ and Microsoft APIs, 62,076 users remained in our dataset.  
Then, to validate our method, we applied textual pattern recognition algorithms to parse a list of patterns that specifically describe users’ age in  self-descriptions (e.g., ``I am 27 years old'', ``I'm 23''), adopting Jang\etal~\cite{jang2015generation}. Young Adults are over-represented as compared to US Census Data, but this is in line with increased use of Online Social Networks by younger users\footnote{e.g., see \url{https://goo.gl/3aCrVw}}.
\subsection{Representativeness of Data}
Although we mainly analyze the data from {\fb}, {\ing}, {\twt} and {\lnkd}, {\me} provided the indispensable link between the identities of a single user across the other platforms. {\me} was chosen for this study as it was the largest such site at the time of crawling\footnote{When the data was collected in Aug 2015, {\me} was ranked in the top 3000 of all websites in the Alexa rankings, and although it is ranked $\approx 5000$ now, it still remains the largest such website.}, and thus provided the largest breadth and coverage across the major social networks of interest. 
However, the  above differences from `expected' demographics in terms of age and gender highlight the issue of `representativeness bias'~\cite{Tufekci14} introduced by the dataset. Apart from the gender and age bias noted above, according to Alexa siteinfo, {\me} also has a more than expected  number of visitors with graduate school eduction. 

However, the notion of `expected' can itself  be problematic: There can be more than one reasonable baseline -- e.g., our data could be compared against the US Census data since our data is primarily from US cities; or it could be compared against all users of social networks. More importantly, Alexa siteinfo indicates that there are differences amongst social networking sites in general -- {\lnkd} is used disproportionately more by those with graduate school background, while {\twt} and {\fb} have slightly more than average number of visitors with an educational background of 'college' and 'some college' respectively, and {\ing} user base is more female than male. This raises the interesting question of what `representative' means for a multi-platform study such as ours: A user who has an account on  {\ing} may not be representative of users on {\lnkd} and vice-versa. Indeed, users who actively use (or have accounts on) multiple social networks are likely not representative of the `average' social network user -- thus a user who has an account on both {\ing} and {\lnkd} is unlikely to be a representative user of either platform, or of an entirely different platform such as {\fb}. 

Yet, given the prevalence of and recent increase in the use of multiple social networks~\cite{duggan2016social,zhong2014social}, we expect that effects such as the ones we see here may become more and more typical, even if it is not typical usage today, and therefore believe our results are a valuable insight into how users shape their online personae across social network platforms. Follow-on work may be required to confirm the wider applicability of our results, and to better understand whether these kinds of bias may be an unavoidable issue for any multi-platform study, or if there are ways to overcome some of them.

\begin{figure*}[tb]
\small
\centering
\subfloat[Face number]{\includegraphics[width=0.33\textwidth]{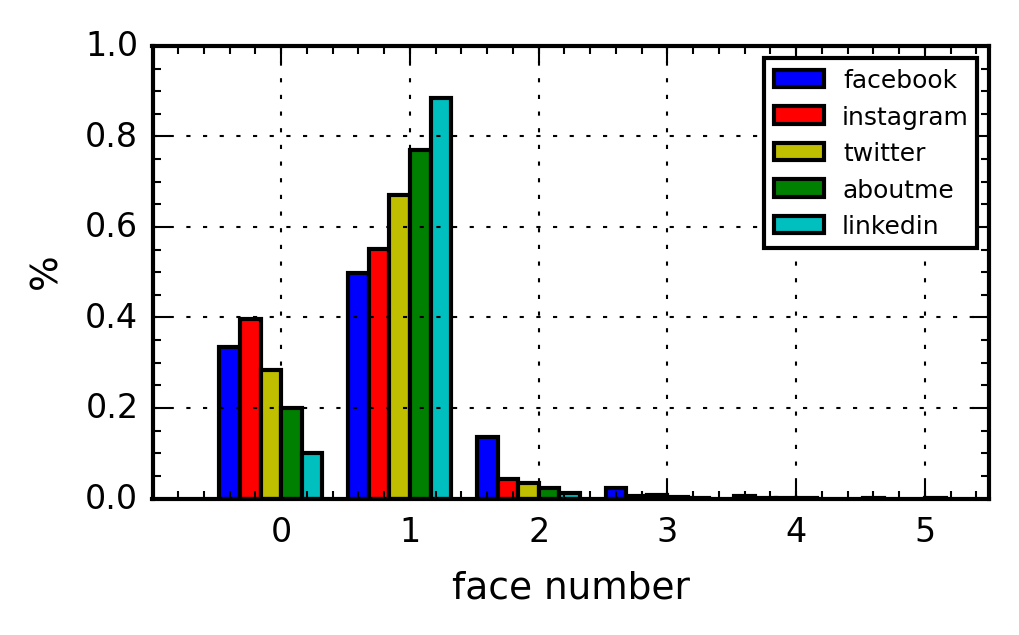}\label{fig:measurement-face-number-pdf}}
\subfloat[Smiling score]{\includegraphics[width=0.33\textwidth]{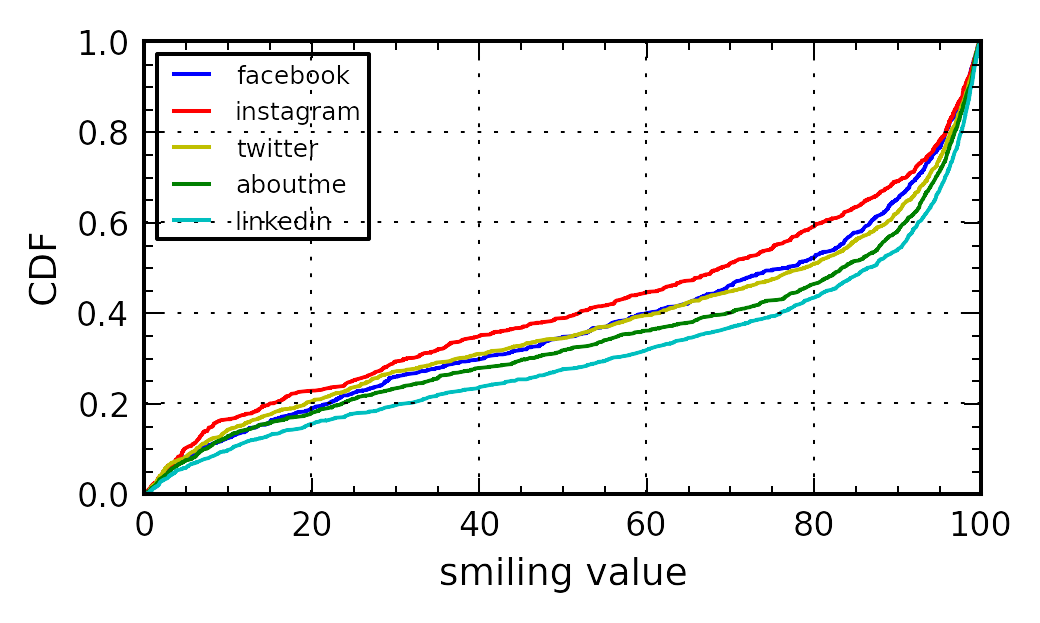}\label{fig:measurement-face-smiling-cdf}}
\subfloat[Image category]{\includegraphics[width=0.33\textwidth]{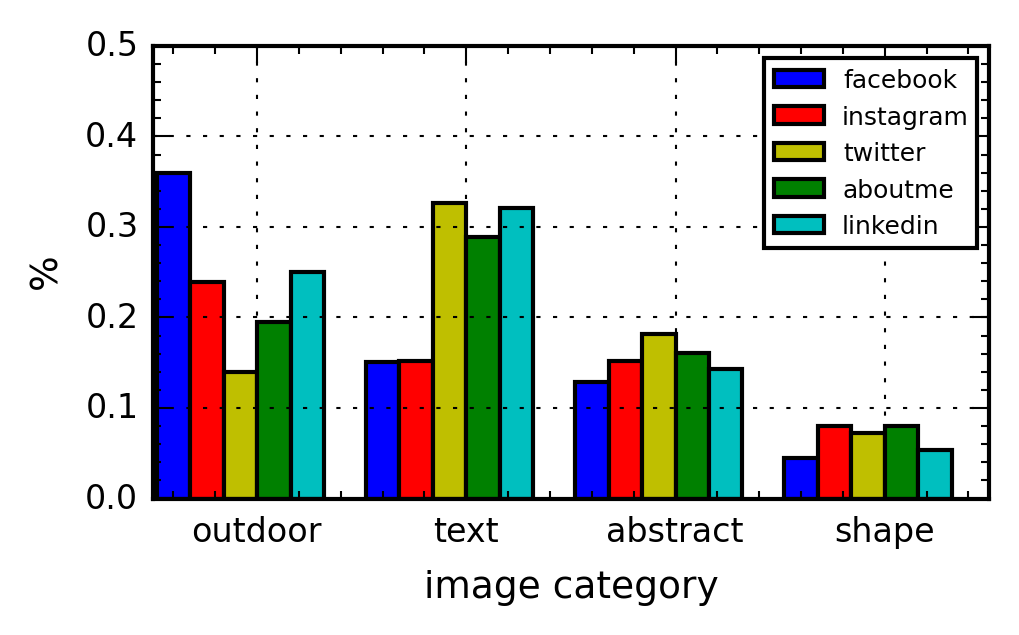}\label{fig:measurement-img-cate}}
\caption{\textbf{The distribution of the detected (a) face number, (b) smiling score, (c) image category of users' profile images at different social networks.} Note that we only consider images with 1 face in (b) and images not in the ``people'' category in (c).}
\label{fig:measurement-dis}
	\vspace{-0.5cm}
\end{figure*}

\section{RQ1 - Different Norms on Different Networks?}
\label{sec-3}

In the following, we aim to answer RQ1 and quantify differences between the ways {\me} users present themselves on different social platforms. Informed by Goffman's self-representation theory~\cite{goffman1959presentation}, we search for both  explicit verbal self-expressions they attach to their profile texts as well as implicit hidden cues they encode (whether intentionally or not) in their profiles images. For instance, we examine whether a profile picture features friend(s), sunglasses or an outdoor landscape, or whether it contains a smiling portrait picture. We interpret these as indicators of  norms of formality or informality -- or alternately, the extent to which a user is managing her public impression\footnote{To recognise people and faces, we have experimented with both Face++ and Microsoft APIs for detecting the number of faces and have seen a high level of consistency between two. However, the results are similar, and we only present the results from Face++. In contrast, our analyses on smiling scores and glasses are based only on Face++ API, while image category results only on Microsoft API, since these features were only available in one of the libraries.}.

\subtitle{Face Number.}
In \fref{fig:measurement-face-number-pdf}, we analyze the number of faces detected on the profile images across multiple social networks. Note that on {\lnkd}, presumably a more formal social network, the majority (90\%) of profile images are portraits of a single person, while  {\fb} and {\ing}, presumably more informal social networks, less than 60\% of images are portraits of a single person. 
Similarly, both {\twt} and {\me} have a higher proportion of profiles that have portraits than   {\fb} and {\ing}. The latter two platforms appear to surface an interesting convention: a significant minority (up to 40\%) of users who use a non-face image (e.g., cartoon, outdoor landscapes, etc.) as their profile picture. 
In alignment with this convention, {\fb} has also a non-trivial share  (15\%) of group portraits with more than one face (unlike {\lnkd} profiles where this share is less than 1\%), which may be attributed to the emphasis that {\fb} users place on intimate social relationships in their profiles~\cite{mod2010reading,strano2008user}.

\subtitle{Smiling Score.}
Next, we detect the extent to which users are smiling on their profile images by measuring the so-called {\em smiling score}, ranging from $0$ (i.e., no smile) to $100$ (i.e., laugh) as provided by Face++ API. Note that only profiles with 1 face are considered here. \fref{fig:measurement-face-smiling-cdf} presents the distribution of smiling scores in user profile images for 5 social networks. The profile images on more formal platforms (e.g., {\lnkd}) tend to have higher smiling scores than those on informal ones (e.g., {\ing} and {\fb}). We take this is an indication that users tend to manage their professional impression on more formal social platforms -- such as {\lnkd} than they do in informal ones -- such as {\fb}.

\subtitle{Eyeglasses.} 
From Face++ API, we could identify two types of eyeglasses: normal eyeglasses 
and sunglasses. 
For normal eyeglasses, we find a consistent trend across all social networks with about 17\% wearing them. 
But sunglasses, which usually are related to leisure activities, are less likely to appear in profile images from formal platforms (e.g., 2.1\% in {\lnkd}), compared with those from informal social platforms (e.g., 7.3\% in {\fb} and 7.2\% in {\ing}).

\subtitle{Image Category.} Finally, we examine the categories of profile images detected by the Microsoft API\footnote{The taxonomy of image categories can be found at \url{https://www.projectoxford.ai/doc/vision/visual-features}.}. To this end, we exclude all portraits from our analysis ($\approx$  50\% of all profile images) and analyze the image categories of all remaining profile images. \fref{fig:measurement-img-cate} presents the distribution of the top-4 image categories, namely, \emph{outdoor}, \emph{text}, \emph{abstract}, and \emph{shape}, each of which accounts for at least 5\% of non-facial images in every social network. We observe that {\fb} and {\ing} users tend to use more of \emph{outdoor} images on their profiles. This can be explained by the emphasis on non-professional activities -- such as travel experience -- in communicating with their  peers on these networks~\cite{sharda2009tourism}. In contrast, \emph{text}, \emph{abstract}, and \emph{shape} images are more dominating among {\twt} and {\me} profiles. A non-exhaustive manual inspection suggests that this is, in some cases at least, an instance of expressing themselves as recognisable brands.

\begin{figure} 
\centering
\subfloat[Lengths of self-description]{%
\hspace{-0.2in}
\includegraphics[width=0.26\textwidth]{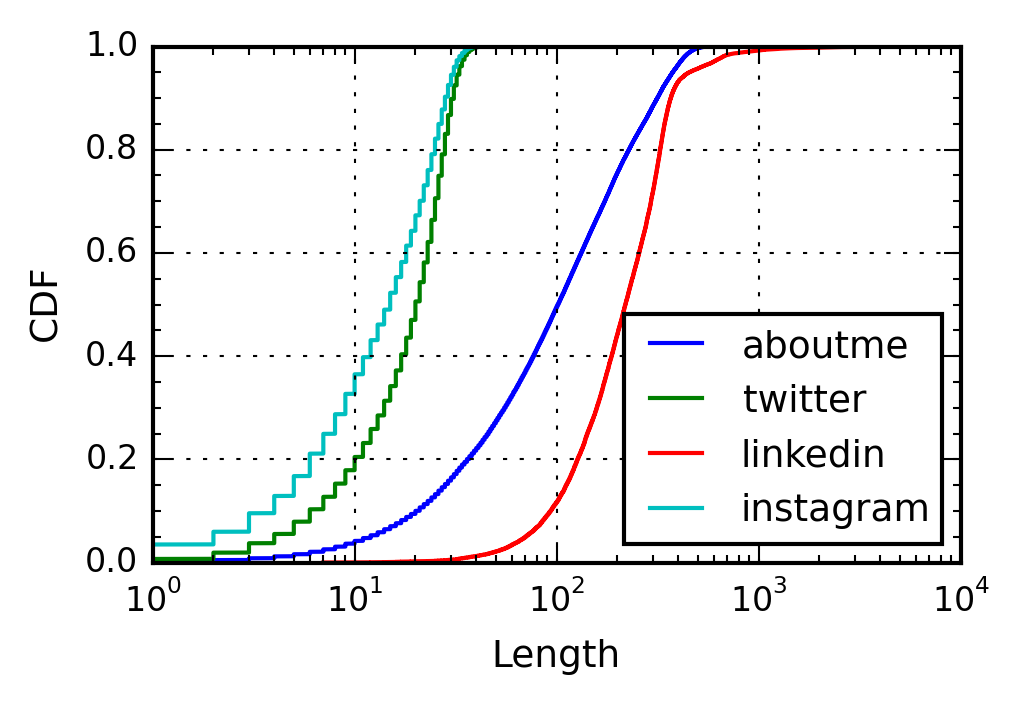}
\label{fig:pf_len}
}
\subfloat[Similarity ({\me})]{%
\hspace{-0.1in}
\includegraphics[width=0.26\textwidth]{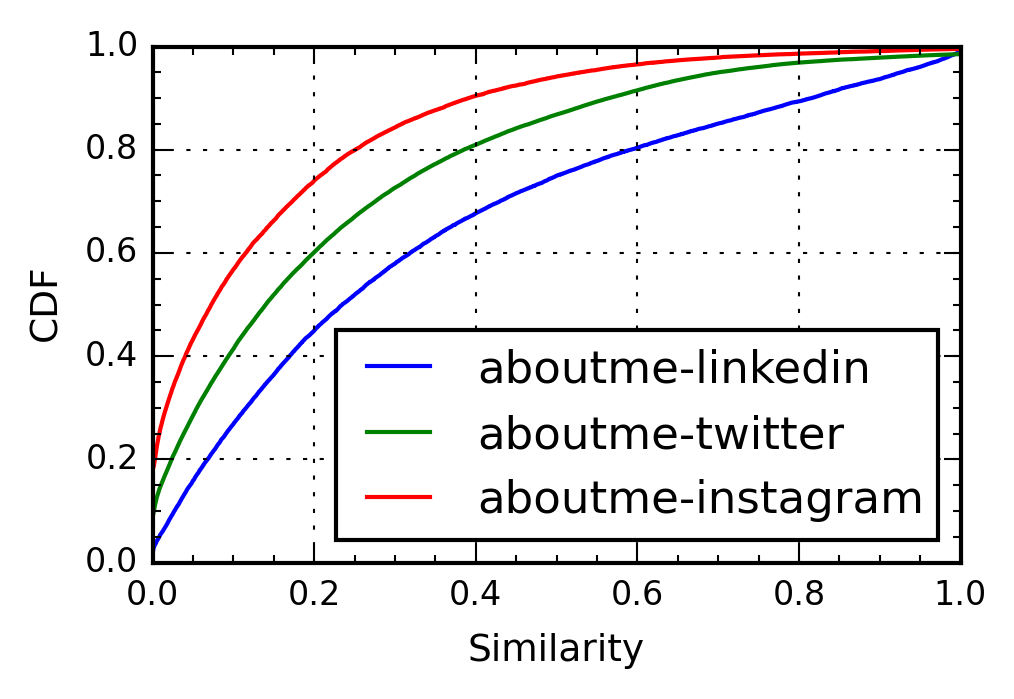}%
\label{fig:pf_sim_me}%
}

	\caption{\textbf{Difference of self-descriptions}: (a) Distribution of lengths of self-description and (b) pair-wise TF-IDF similarity scores across different social networks.}
\label{fig:pf_sim}
	\vspace{-0.5cm}
\end{figure}

\subtitle{Differences in Self-descriptions.} Due to the variation of functionality and specific limitations of social networks, a user might tailor her profiles differently for a given platform. 
As \fref{fig:pf_len} shows the distributions on the length (i.e., the number of words) of self-descriptions across platforms, note that both {\me} and {\lnkd} are skewed to the right, indicating that most of self-descriptions are relatively long. However,  the distributions of {\twt} and {\ing} are somewhat irregular, probably due to the artificially-set length limitation. 

Next, for each possible pair of social network profiles of a user, we  compute the TF-IDF similarities between the self-descriptions\footnote{Recall that there are no self-descriptions from Facebook.}.  The features are normalized to reduce the impact of profile length. The similarity score is between 0 (i.e., very different) and 1 (i.e., exact match). For instance, demonstrated in \fref{fig:pf_sim_me} is the CDF of similarity scores for  {\me}. As expected, the self-description of {\me} is most similar to that of {\lnkd} for the same user, presumably due to the relatively similar functions of both social networks. 

\begin{figure*} 
\centering
\subfloat[{\me}]{%
\includegraphics[width=0.5\columnwidth]{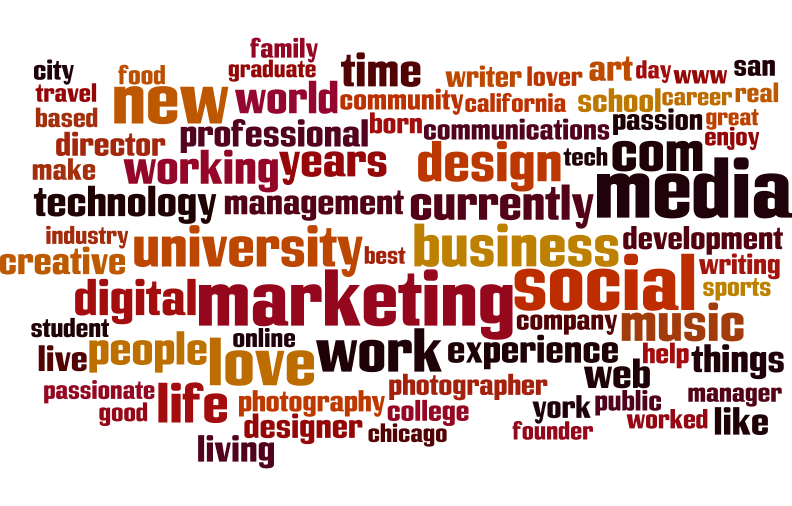}%
\label{fig:wc_me}%
}
\subfloat[{\lnkd}]{%
\includegraphics[width=0.5\columnwidth]{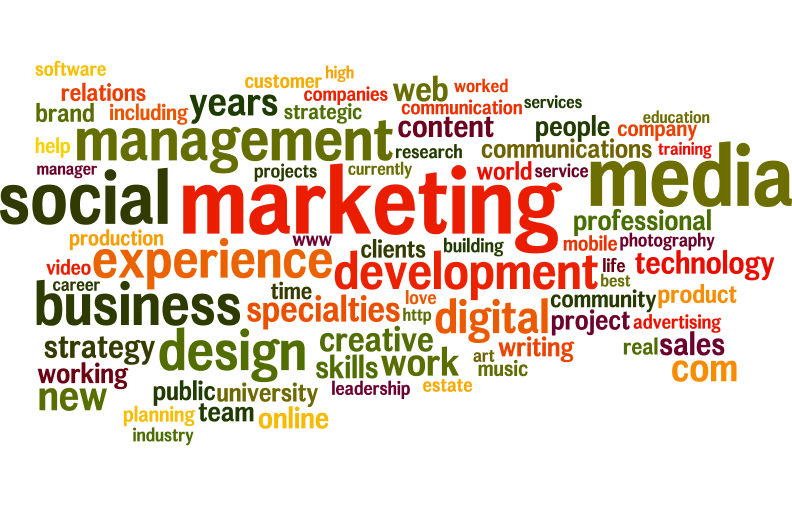}%
\label{fig:wc_in}%
}
\subfloat[{\twt}]{%
\includegraphics[width=0.5\columnwidth]{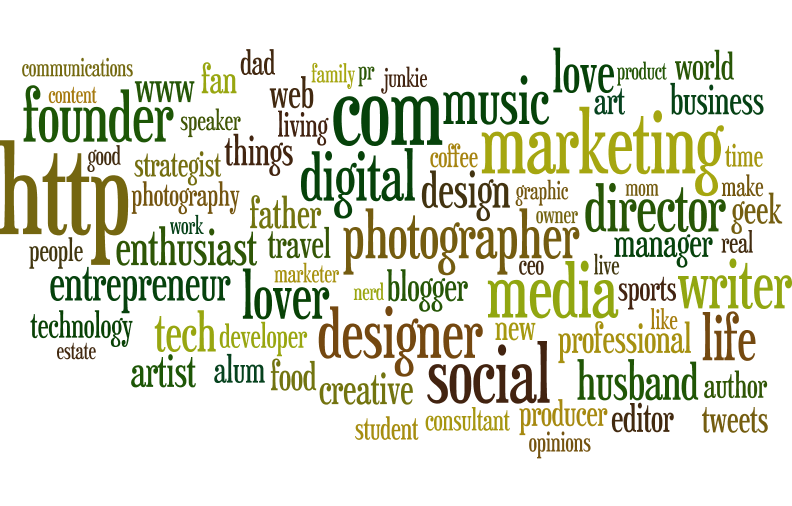}%
\label{fig:wc_tw}%
} 
\subfloat[{\ing}]{%
\includegraphics[width=0.5\columnwidth]{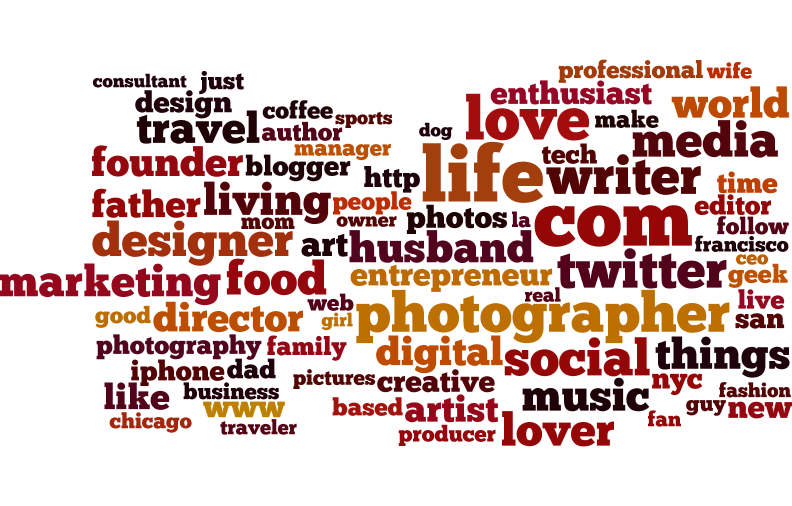}%
\label{fig:wc_ig}%
}
\caption{\textbf{Word clouds made from self-descriptions.} 
}
\label{fig:wc}
\end{figure*}
\subtitle{Word Clouds.}
As discussed in previous sections, each social network tends to develop its own culture, which results in a variation of profiles even for the same user. 
To  summarize the  themes of each social network, in \fref{fig:wc}, we visualize self-descriptions in profiles as  word clouds, generated by a Python library\footnote{\url{https://github.com/amueller/word_cloud}}. It is interesting to see that {\lnkd} and {\me} self-descriptions  concern topics around employments, including professional terminologies (e.g., development, project, experience), types of industry (e.g., marketing, media, business, management), location (e.g., ``new'' for location names like New York and New Jersey) and experiences (e.g., ``year''). However, {\ing} self-descriptions show more relaxed roles of users such as  ``life'', ``love'', ``lover'', ``food'', ``music'', and ``travel''. On the other hand, {\twt} self-descriptions  are somewhat a mix between two groups, heavily comprised  of words such as ``love'', ``marketing', ``write'', and ``social''.

\subtitle{Discussion.} 
The above results suggest that there is a spectrum of norms across the social platforms, with a consistent difference between the more \emph{professional} networks such as {\lnkd} and the more \emph{informal} networks like {\ing}~\cite{van2013you}. Many if not all of the above differences can be understood in terms of the professional or non-professional nature of the platform. 

The fact that a single user may have more than one kind of profile is explained by the faceted identity theory~\cite{farnham2011faceted}, which posits that  people have multi-faceted identities, and enable different aspects of their personalities depending on the social context.
Users may be more focused on managing public impression~\cite{roberts2005changing} in professional circles, but may not be when with friends and family.
The strength of the boundaries between these different social roles may vary across individuals~\cite{clark2000work,farnham2011faceted} and may impact the patterns of daily routines~\cite{ashforth2000all} to different extents for different individuals.

More interestingly, there are empirical evidences~\cite{ozenc2011life} suggesting that, to manage communication within different social roles, people strategically employ different social media channels. For instance, on social networks focused on building and maintaining external professional networks (e.g., {\lnkd}) -- users may want to present themselves in a formal and professional way -- whereas on general-purpose social networks (e.g., {\fb}), users may keep relationships more informal and  adjust their profiles accordingly~\cite{skeels2009social}.

\section{RQ2 - Profile Classification}
\label{sec-ml1}

Building on the analysis from RQ1, now, we ask whether users fit in to the norms of those social networks and attempt to answer this indirectly by studying how accurately one can predict profiles in different social networks.
Note that we build a non user-specific model to see whether users fit in to a given social network in a consistent manner .
We formulate the \emph{profile classification} problem as follows:
\vspace{0.1in}
\begin{myPro}[Profile Classification]
\emph{
Given a profile image (or self-description) of a  user $u$, predict whether it \emph{fits in} the profile conventions of a given social network $n$.
}
\end{myPro}
We tackle this problem using a supervised learning approach where we exploit profile images or self-descriptions originated from different social networks and train a classification model to predict the social networks that they have been picked from.

For each profile image  in our dataset, we extract the aforementioned facial and deep learning features,
including face number, gender, age, race, smiling, glasses, and face pose. We also consider 5,096 features extracted using convolution neural networks including a 1,000 dimensional vector of recognized objects and a 4,096 dimensional vector extracted from the pre-trained deep neural network. Similarly, we also extract textual features from self-descriptions and construct a Word2Vec vector. 
For the purpose of this analysis, we validate the models using a Random Decision Forest classifier\footnote{We used a Random Forest implementation from the SKLearn package with $\sqrt{n_{features}}$ split and 100 estimators (other values from 10 to 1000 were also tested, but 100 showed the best trade-off between speed and prediction performance). $\chi^2$ feature selection is applied to select 1000 most relevant features for the classification with profile images.}  and report the results of the 10-fold cross-validation.

\begin{table}
{\small
\centering
{
\subfloat[Using profile images]{
\begin{tabular}{c|ccccc}
	\hline 
	&\tabhead{ACC} & \tabhead{P} & \tabhead{R} & \tabhead{F} & \tabhead{AUC} \\ 
	\hline 
	{\ing} & 0.829 & 0.790 & 0.896 & 0.840 & 0.829 \\
	{\fb}& 0.770 & 0.746 & 0.824 & 0.782 & 0.769 \\
	{\me} & 0.730 & 0.723 & 0.748 & 0.736 & 0.730 \\
	{\lnkd} & 0.687 & 0.698 & 0.665 & 0.681 & 0.687 \\
	{\twt} & 0.657 & 0.644 & 0.707 & 0.673 & 0.657\\
	\hline 
\end{tabular}
\label{tab:ml1-1vsother-image}
}

\subfloat[Using self-description]{
\begin{tabular}{c|ccccc}
	\hline 
	&\tabhead{ACC} & \tabhead{P} & \tabhead{R} & \tabhead{F} & \tabhead{AUC} \\ 
	\hline 
	{\ing}& 0.768 & 0.571 & 0.288 & 0.383 & 0.608 \\
	{\me} & 0.866 & 0.765 & 0.643 & 0.716 & 0.802 \\
	{\lnkd} & 0.871 & 0.720 & 0.797 & 0.756 & 0.847 \\
	{\twt} & 0.788 & 0.598 & 0.468 & 0.525 & 0.681\\
	\hline 
\end{tabular}
\label{tab:ml1-1vsother-text}
}
\caption{\textbf{Performance of the \emph{one-vs-others} profile classification.} We evaluate the performance using Accuracy (ACC), Precision (P), Recall (R), F-score (F), and Area Under the receiver-operator characteristic Curve (AUC). 
}
\label{tab:ml1-1vsother}
}
}
	\vspace{-0.5cm}
\end{table}

\subtitle{One-vs-others Classification.} 
To start with, we train a set of one-vs-others classifiers (one for each social network) which for a given image or self-description are trained to distinguish between social networks that they belong to. To this end, for each classifier, we label profile images/self-descriptions picked from the corresponding social network as positive instances and randomly sample the same number of profile images/self-descriptions from the other four social networks (i.e., ``others'') as negative instances. 
The results of the one-vs-others experiments using profile images and self-descriptions are summarized in \tref{tab:ml1-1vsother}.

First, we note a high prediction performance for one-vs-others profile classification problem--e.g., high AUC scores  of up to 0.829 for {\ing} profile images and up to 0.847 for {\lnkd} self-descriptions. This suggests that profile conventions of individual social networks can be successfully recognised by machines with high accuracies. However, prediction performance varies a lot across  social networks. For example, the AUC of classifying {\twt} profiles is only 0.657 in comparison with the best-in-class AUC of 0.829 of {\ing}. In general, we observe that profile conventions in informal social networks (e.g., {\fb} and {\ing}) are much more recognisable using profile images than in professional social networks (e.g., {\lnkd} and {\me}). In contrast, self-descriptions in professional social networks perform much better than in informal social networks. 
This implies that the conventions of professional social networks are mainly expressed through users' self-descriptions, while profile image is a channel for expression on informal social networks.

\subtitle{One-vs-one Classification.} 
In the \emph{one-vs-one} classification, we aim to identify profile images between pairs of social networks: i.e., given profiles from {\em two} different social networks, we train a binary classifier to distinguish an origin social network for each given profile. The results of the experiments are summarized in \tref{tab:ml1-1vs1}. We note that the results highlight two distinct groups, columns  in \tref{tab:ml1-1vs1} separated by dotted lines, among considered social networks--{\ing} and {\fb}, on the one hand, and {\me} and {\lnkd}, on the other hand, with {\twt} in between two groups. In general, intra-group prediction is low while inter-group prediction is high. For instance, AUC for the {\ing}-{\fb} pair is 0.744, while that for the {\ing}-{\me} pair is 0.905.

In other words, this result highlights that the conventions inside each of the two groups -- professional networks (i.e., {\me}, {\lnkd} and to some extent {\twt}) on the one hand, and the more informal networks on the other hand (i.e., {\ing} and {\fb}) are very similar. This intra-group similarity  makes it difficult to distinguish between profile images randomly picked from two.  Similarly, it is much easier to distinguish between a profile image taken from a professional social network and an informal one, suggesting that conventions among the two groups of networks are very different. We note that this result resonates with our findings from  previous sections.

\begin{table}
{\small
\centering
{
\subfloat[Using profile images]{
\begin{tabular}{c|cc:cc}
	\hline 
	&\tabhead{{\ing}} & \tabhead{{\fb}} & \tabhead{{\me}} & \tabhead{{\lnkd}}\\ 
	\hline 
	\tabhead{{\fb}} & {0.744} &  & &    \\
	\hdashline
	\tabhead{{\me}} & 0.905 & 0.873 & &   \\
	\tabhead{{\lnkd}} & 0.890 & 0.834 & {0.691} & \\
	\tabhead{{\twt}} & 0.879 & 0.835 & {0.571} & {0.599}\\
	\hline 
\end{tabular}
\label{tab:ml1-1vs1-image}
}\\
\subfloat[Using self-description]{
\begin{tabular}{c|c:cc}
	\hline 
	&\tabhead{{\ing}} & \tabhead{{\me}} & \tabhead{{\lnkd}}\\ 
	\hline 
	\tabhead{{\me}} & 0.894 & &   \\
	\tabhead{{\lnkd}} & 0.863 & 0.869  & \\
	\hdashline
	\tabhead{{\twt}} & 0.665 & 0.867 & 0.974\\
	\hline 
\end{tabular}
\label{tab:ml1-1vs1-text}
}

\subfloat[Using combined features]{
\begin{tabular}{c|c:cc}
	\hline 
	&\tabhead{{\ing}} & \tabhead{{\me}} & \tabhead{{\lnkd}}\\ 
	\hline 
	\tabhead{{\me}} & 0.952 & &   \\
	\tabhead{{\lnkd}} & 0.904 & 0.756  & \\
	\hdashline
	\tabhead{{\twt}} & 0.880 & 0.862 & 0.639\\
	\hline 
\end{tabular}
\label{tab:ml1-1vs1-combined}
}
\caption{\textbf{The \emph{One-vs-one} performance (AUC) of profile classification.} 
Note the emergence of two groups, separated by dotted lines. Intra-group prediction is low while inter-group prediction is high.}
\label{tab:ml1-1vs1}
}}
	\vspace{-0.5cm}
\end{table}

In \tref{tab:ml1-1vs1-combined}, finally, we use the combined hybrid features to \emph{one-vs-one} experiments. It  shows that the combined model takes the advantages of both profile images and self-descriptions, and improves  most pairwise predictions from~\tref{tab:ml1-1vs1-text}, with the exceptions of {\me}-{\lnkd} and {\twt}-{\lnkd} pairs.

\subtitle{Summary.} In this section, we demonstrated that using either profile images or self-descriptions, it is indeed possible to accurately predict profiles in different social networks. Indeed, it shows that most of users tend to fit in (by means of profiles) to the conventions of a particular social network.
In addition, the proposed profile classification problem has a practical ramification. For instance, by turning the classification problem into the recommendation problem, one can build a tool to recommend the most appropriate profile image (among many choices) on a particular social network site for a given user.

\begin{figure*}[tb]
	\small
\centering
\subfloat[Smiling (gender)]{\includegraphics[width=0.32\textwidth]{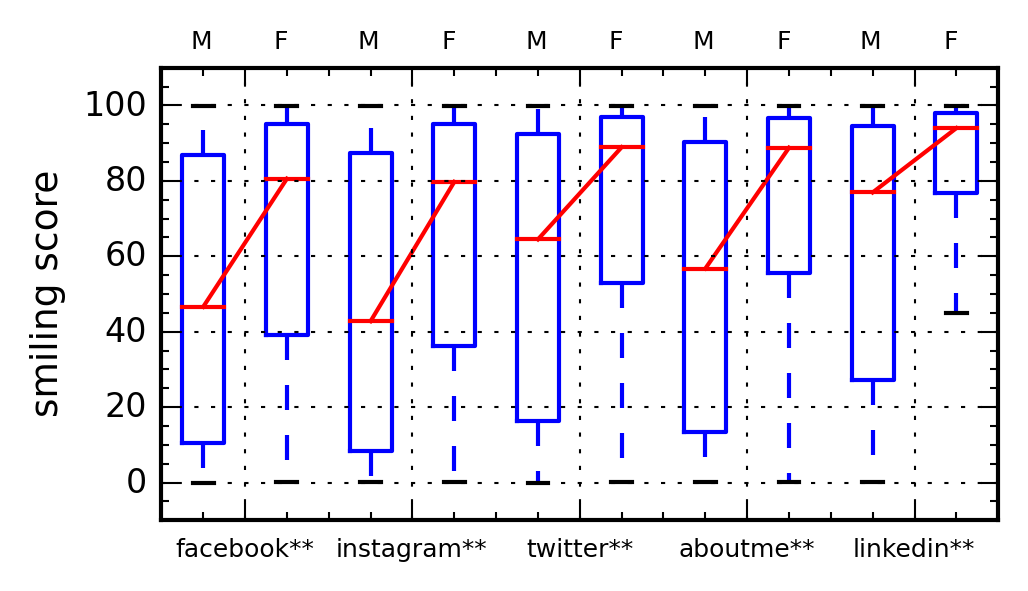}\label{fig:demographic-smiling-gender}}
\subfloat[Eyeglasses (gender)]{\includegraphics[width=0.32\textwidth]{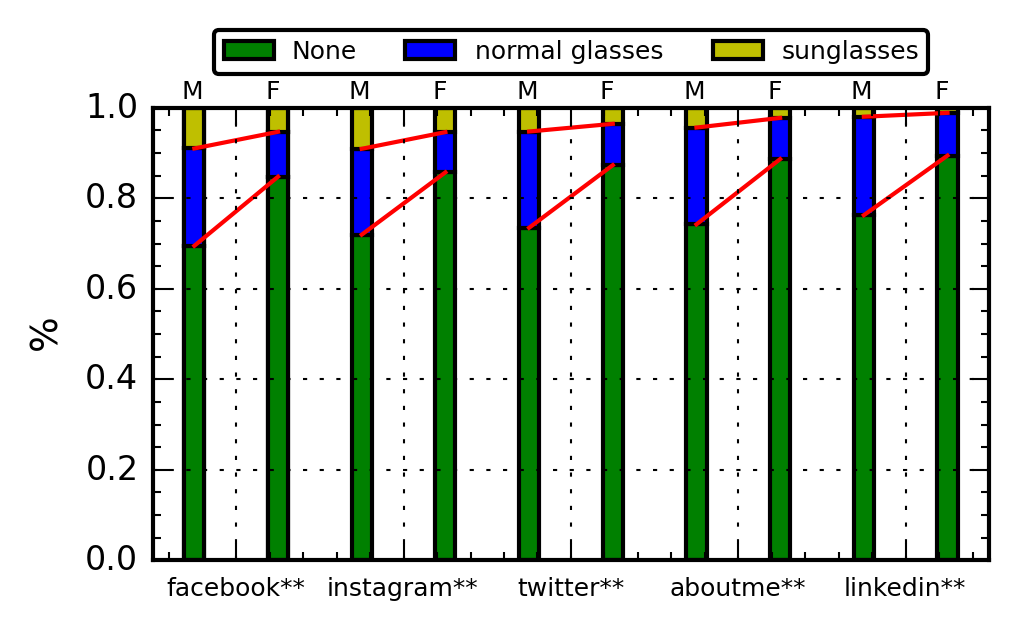}\label{fig:demographic-glasses-gender}}
\subfloat[Partners (gender)]{\includegraphics[width=0.34\textwidth]{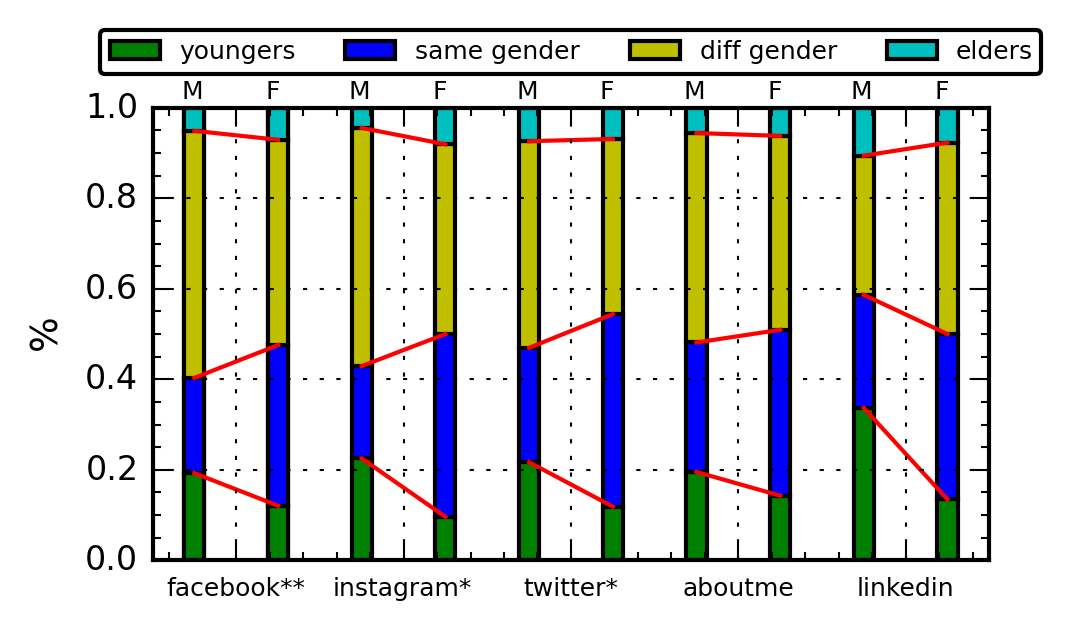}\label{fig:demographic-partner-gender}}
\\
\subfloat[Smiling (age)]{\includegraphics[width=0.32\textwidth]{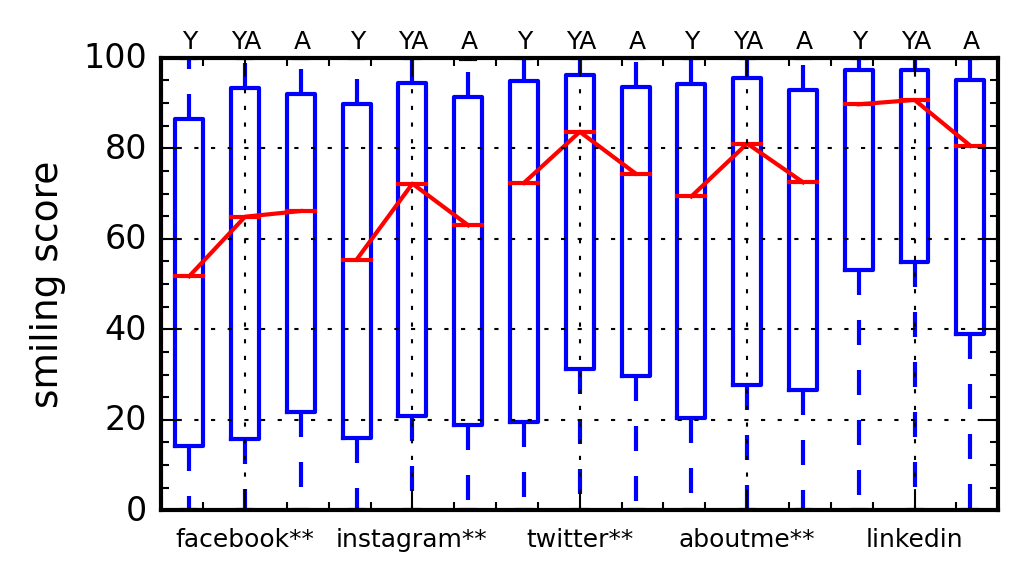}\label{fig:demographic-smiling-age}}
\subfloat[Eyeglasses (age)]{\includegraphics[width=0.32\textwidth]{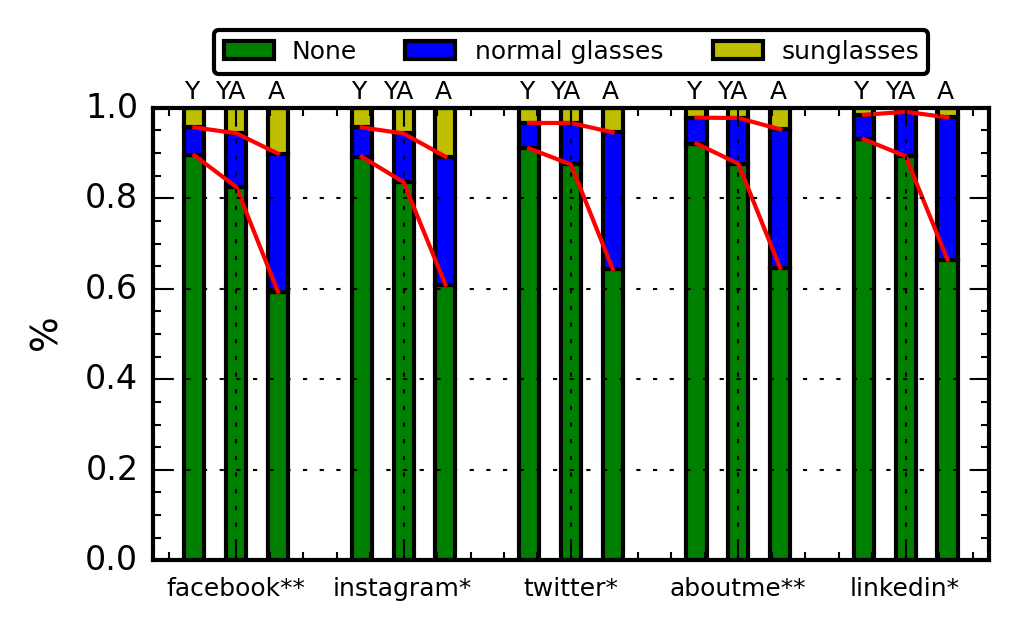}\label{fig:demographic-glasses-age}}
\subfloat[Partners (age)]{\includegraphics[width=0.34\textwidth]{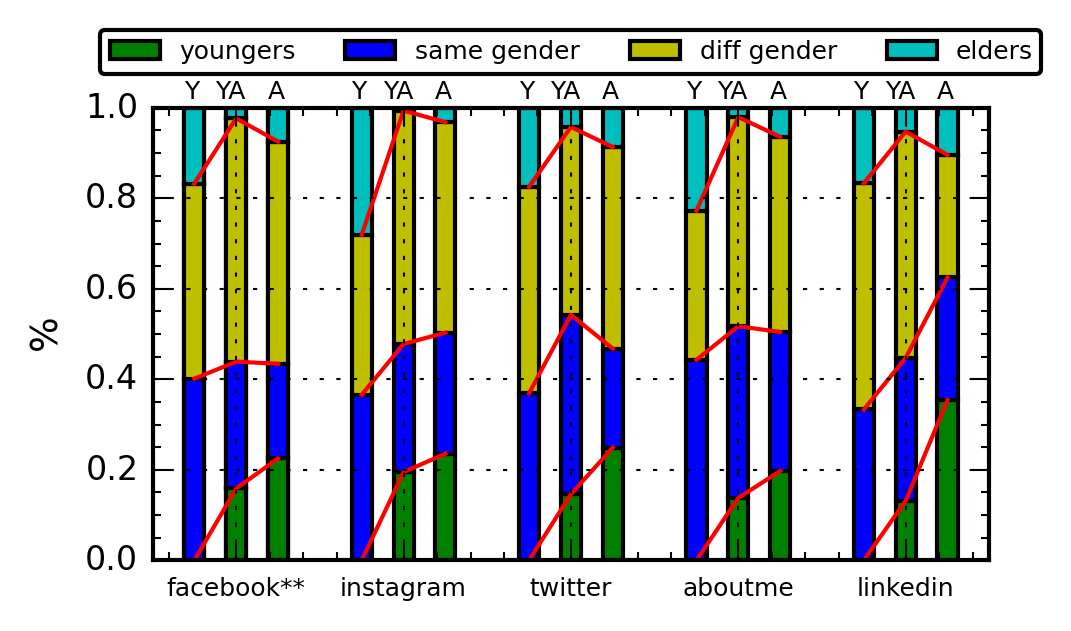}\label{fig:demographic-partner-age}}
\caption{\textbf{The comparison of the facial features with different genders and ages.} For gender analyzes (a-c), ``M'' represents results of males, and ``F'' is for females. For age analyzes (d-f), ``Y'' is for youth, ``YA'' is for young adults and ``A'' is for adults. We examine the differences among distributions for different genders/ages using Kolmogorov-Smirnov test, and label ``**'' for cases with p-value $p<0.005$, and ``*''  for cases with $p<0.05$.}
\label{fig:demographic-img}
	\vspace{-0.5cm}
\end{figure*}

\section{RQ3 - Gender and Age Differences?}

In the previous two sections, we have discussed conventions found in user profiles across multiple social networks and analyzed the extent to which user profiles \emph{fit in} the conventions of individual social networks. Building on these analysis, in the current section, we further investigate the differences between the way users from different demographic groups express themselves in their social media profiles. To this end, we divide users into different groups according to their gender and age information and analyze the differences in profile images and self-descriptions across different groups. In particular, we examine the discrepancies in aspects such as 
\emph{smiling score}, \emph{eyeglasses} and \emph{partners} as identified by the facial recognition libraries.

\subtitle{Smiling Scores.} We first examine emotional expression in profile faces. To do so, we focus on profile images with a single face only, and compare the smiling scores across different demographic groups. In \fref{fig:demographic-smiling-gender}, we compare the distributions of smiling scores for users with different genders. We note that women tend to smile more in the profile images across all five social networks (with $p<0.005$).  This result is consistent with the several previous studies in psychology~\cite{coats1996gender,mcclure2000meta,strano2008user,tifferet2014gender}. Indeed, according to Coats and Feldmen~\cite{coats1996gender}, women who display positive emotional expressions tend to be rated of a higher social status, while men who do so risk being rated as having low social status; thus there is 
motivation for women to fit in by  smiling, and for men not to.

We also compare the smiling score for users in different age groups in \fref{fig:demographic-smiling-age}. On the one hand, we find that users in adult group (A, age $\ge35$) tend to have lower smiling scores than young adult group (YA, age in $26-34$) ($p<0.005$, except {\lnkd}), which is consistent with existing psychological theories~\cite{folster2014facial,gross1997emotion} that older people tend to appear less expressive. On the other hand, surprisingly, we find that users from the youths group (Y, age $\le 25$) also have lower smiling score. We suspect this is due to higher popularity of novel image categories such as selfies amongst youth, where smiles are less common or less pronounced~\cite{souza15dawn}.

\subtitle{Eyeglasses.} We also examine eyeglasses people wear in profile images. As discussed before, we identify two types of eyeglasses: normal glasses and sunglasses. 
Normal glasses are mainly used for vision correction.  In \fref{fig:demographic-glasses-gender} (blue parts), we compare the usage of normal glasses for different genders. Interestingly, we see that more males wear normal glasses than females (with $p<0.005$), especially for platforms like {\lnkd}. This result is in contrast with the findings of the U.S. National Eye Institute (NEI)~\cite{nei-stats} on myopia, the most cause for corrective eye lenses~\cite{nei-facts}: the NEI report finds that women have higher myopia rate than men. A potential 
explanation is that wearing normal glasses is considered more intelligent and formal~\cite{edwards1987effects}. We see more normal glasses for males, because males are thought to dress more formally than females~\cite{chawla1992impact,sebastian2008formal}. An alternative explanation can be that there is a social pressure on women not to wear glasses, which may influence their choice of profile image. Both explanations, however, are indicative of a choice towards fitting in with expected gender-specific norms or social pressures.

\subtitle{Partners.} So far, we have studied how users look like in their profile images. We next explore the partners that appear in users' profile images. We focus on profile images with 2 faces\footnote{Due to the data limit, as shown in \fref{fig:demographic-partner-gender} and \ref{fig:demographic-partner-age}, the results of partners are only statistically  significant for {\fb} with $p<0.005$, although similar trends can also observed in other networks.}. Since for each user we have several images, we could easily identify the user ($u$) and the partner ($p$) from 2 faces by matching gender and age. Then we decide the relationship between them. To do this, we check the age differences between $u$ and $p$ with a threshold $X$\footnote{We present results with $X=20$, although similar results can be observed for $X=\left\{ 10, 15, 25, 30\right\}$}. If $|age(u)-age(p)|>X$, we consider they are in different generations, otherwise in the same generation. When $u$ and $p$ are in different generations, we compare their ages again, if $age(u) > age(p)$, we say the image is with ``younger'', otherwise, we say with ``elder''. When $u$ and $p$ are in same generation, we compare their genders, if $gender(u)$ and $gender(p)$ are the same, we say the profile image is with ``same gender'' friends, otherwise, we say it is with ``different gender'' friends. 

Using this framework, we compare partners for users in different groups. In \fref{fig:demographic-partner-age}, we find that with  increasing  age, users tend to have more profile images with ``youngers'', which is likely be their kids, and less images with ``same gender'' friends. For ``elders'', or parents in most cases, users in $26-34$ group have more images then users in $\le25$ and $\ge35$ groups.
Surprisingly, in \fref{fig:demographic-partner-gender}, we observe males have more images with ``youngers'' (likely, kids) and ``different gender'' friends than females. But females tend to  use more images with 
``same gender'' friends.  This result is consistent with \cite{strano2008user}, which finds that females are more likely to emphasize friendships in their profile images than males.

\subtitle{Summary and discussion} In this section, we have answered RQ3, showing that although users are adapting to social norms of a given platform, there are distinct differences in the way that different genders and age groups present themselves. For example, females smile more than males and tend to avoid the use of spectacles in their profile pictures. Although these differences are interesting in and of themselves, it is more interesting that there are statistically significant and consistent differences across the population, indicating that profile construction is informed by gender and age demographics. Gender and age are the most basic social groups.  It would be interesting to consider whether other more sophisticated  groupings of people (e.g., by party affiliation) tend to construct their profiles in similar ways within the group.

\section{Discussion, Limitations, and Conclusion}
\label{sec:conclusions}

In this paper, we used the personal web pages of 116,998 users from {\me} to identify and extract a matched set of  user profiles on four major social networks ({\fb}, {\twt}, {ing} and {\lnkd}), and studied the following three questions: (1) whether users express themselves differently on different social networks, to ``fit into'' the ``culture'' or expected ``norms'' of the network  (2) whether different users ``fit-in'' in a consistent manner, i.e., whether we can learn  a non user-specific model that distinguishes  a profile as fitting or belonging to one social network more than another, and (3) whether there are any universal cross-platform tendencies in how various genders and age groups fit into a given platform.

Answers to these questions can have deep implications for core applications such as advertising or community building: 
The extent to which norms exist as well as the extent to which users (need to) adapt to fit those norms influences what kinds of users are attracted to each  network. This can affect advertising strategies (e.g., brands and advertising styles that find a large audience on say {\lnkd} may not be popular on {\ing}, because the expected norms and core demographic may be different). 
This approach can also facilitate automated generation of stereotype personas - fictitious individuals representing segments of the audience in marketing, design and advertising~\cite{an2017automatic}.

Similarly different strategies may be needed on different platforms for community engagement and growth (e.g., gamification and rewards may promote community engagement on one social network which attracts one kind of user, but not on another network with a different user base that does not like the competitiveness of gamification). If users ``fit in'' in consistent patterns on a platform, then new UI affordances can be developed to help them fit in. At the same time, if users in different demographic groups have different ways of fitting in, or feel pulled to fit in to different extents, then UI affordances need to be sensitive to and support such differences.

Our results indicate that different social networks do have different conventions, and users do fit their different social networks profiles to suit these conventions. Importantly, we confirm that the networks we examine -- {\fb}, {\ing}, {\twt}, {\lnkd} and {\me} -- fall on a spectrum, based on how formal or informal the network's intended purpose is. Profiles on more formal or professional social networks such as {\lnkd} and {\me} are presumably intended to be showcased to an audience of non-friends (e.g., recruiters on {\lnkd}, or random visitors to personal webpages on {\me}) and are therefore constructed differently from profiles on networks such as {\fb} or {\ing}, which appear to be geared more towards friends and relationships. We find these differences both in conventions relating to profile images (e.g., usage of pictures with friends and companions on {\fb} as opposed to formal single-person portraits on {\lnkd}) as well as the choice of text in profile descriptions (\fref{fig:wc}). 

Users' fitting in to conventions are recognisable by simple machine learning classifiers;  networks that are farther apart on the formal / informal spectrum are easier to tell apart from each other. We also find evidence that different genders and age groups fit in to different extents, and that the relative extent of fitting in, in general, is consistent across social network platforms. For instance, women are more likely to have pictures with companions and without glasses. While these differences are important in and of themselves (and we try to explain how these differences arise), in the context of this paper, it is more interesting that such differences exist at all, and that are statistically significant and consistent across the platforms. This allows us to draw conclusions on how social network profiles vary based on how formal or informal the social network platform is.

This paper is intended as a first look at how user profiles vary in the aggregate across different social networks. Although we observe robust effects, as noted before (in the Representativeness of Data subsection), our results may be coloured by the fact that our data is limited to users who have made a profile on {\me}. A second limitation of our study is that it is based solely on data analysis. It would be interesting to validate whether our conclusions  correspond to user motivations in choosing a particular profile image or writing a particular textual self-description. This requires a direct user study, which was out of scope for the current work. 

Our choice of relying solely on data analysis was driven by the observation that even if users \emph{believed} they were not trying to fit in, the data might indicate an \emph{unconscious} bias in user choices towards fitting in with a (potentially subliminally) perceived norm. Thus, we are of the opinion that the data analysis in and of itself can be a valuable first step towards understanding profile construction across social networks. Disambiguating between users' conscious ideas about fitting in and actions observed through a data-based approach would require careful research design and can be the subject of a follow-on work.

\bibliographystyle{aaai}
\sloppy
{
\footnotesize
\setstretch{0.75}
\balance
\bibliography{bib/aboutme} 

\begin{thebibliography}{}

\bibitem[\protect\citeauthoryear{An, Kwak, and Jansen}{2017}]{an2017automatic}
An, J.; Kwak, H.; and Jansen, B.~J.
\newblock 2017.
\newblock Automatic generation of personas using youtube social media data.
\newblock In {\em HICSS}.

\bibitem[\protect\citeauthoryear{Ashforth, Kreiner, and
  Fugate}{2000}]{ashforth2000all}
Ashforth, B.~E.; Kreiner, G.~E.; and Fugate, M.
\newblock 2000.
\newblock All in a day's work: Boundaries and micro role transitions.
\newblock {\em Academy of Management review} 25(3):472--491.

\bibitem[\protect\citeauthoryear{Bakhshi, Shamma, and
  Gilbert}{2014}]{bakhshi2014faces}
Bakhshi, S.; Shamma, D.~A.; and Gilbert, E.
\newblock 2014.
\newblock Faces engage us: Photos with faces attract more likes and comments on
  instagram.
\newblock In {\em ACM CHI},  965--974.

\bibitem[\protect\citeauthoryear{boyd}{2011}]{boyd:2011}
boyd, d.
\newblock 2011.
\newblock Social network sites as networked publics: Affordances, dynamics, and
  implications.
\newblock In {\em Networked Self: Identity, Community, and Culture on Social
  Network Sites}. Routledg.

\bibitem[\protect\citeauthoryear{Chawla, Khan, and
  Cornell}{1992}]{chawla1992impact}
Chawla, S.~K.; Khan, Z.~U.; and Cornell, D.~W.
\newblock 1992.
\newblock The impact of gender and dress on choice of cpa's.
\newblock {\em Journal of Applied Business Research} 8(4):25.

\bibitem[\protect\citeauthoryear{Cimino}{2009}]{cimino2009netiquette}
Cimino, M.
\newblock 2009.
\newblock {\em Netiquette (on-line Etiquette): Tips for Adults \& Teens:
  Facebook, MySpace, Twitter! Terminology--and More}.
\newblock PublishAmerica.

\bibitem[\protect\citeauthoryear{Clark}{2000}]{clark2000work}
Clark, S.~C.
\newblock 2000.
\newblock Work/family border theory: A new theory of work/family balance.
\newblock {\em Human relations} 53(6):747--770.

\bibitem[\protect\citeauthoryear{Coats and Feldman}{1996}]{coats1996gender}
Coats, E.~J., and Feldman, R.~S.
\newblock 1996.
\newblock Gender differences in nonverbal correlates of social status.
\newblock {\em Personality and Social Psychology Bulletin} 22(10):1014--1022.

\bibitem[\protect\citeauthoryear{Deng \bgroup et al\mbox.\egroup
  }{2009}]{deng_imagenet_2009}
Deng, J.; Dong, W.; Socher, R.; Li, L.-J.; Li, K.; and Fei-Fei, L.
\newblock 2009.
\newblock {ImageNet}: A large-scale hierarchical image database.
\newblock In {\em {IEEE} {CVPR}},  248--255.

\bibitem[\protect\citeauthoryear{Donahue \bgroup et al\mbox.\egroup
  }{2013}]{donahue_decaf_2013}
Donahue, J.; Jia, Y.; Vinyals, O.; Hoffman, J.; Zhang, N.; Tzeng, E.; and
  Darrell, T.
\newblock 2013.
\newblock {DeCAF}: A deep convolutional activation feature for generic visual
  recognition.

\bibitem[\protect\citeauthoryear{Edwards}{1987}]{edwards1987effects}
Edwards, K.
\newblock 1987.
\newblock Effects of sex and glasses on attitudes toward intelligence and
  attractiveness.
\newblock {\em Psychological Reports}.

\bibitem[\protect\citeauthoryear{Farnham and
  Churchill}{2011}]{farnham2011faceted}
Farnham, S.~D., and Churchill, E.~F.
\newblock 2011.
\newblock Faceted identity, faceted lives: social and technical issues with
  being yourself online.
\newblock In {\em CSCW},  359--368.
\newblock ACM.

\bibitem[\protect\citeauthoryear{F{\"o}lster, Hess, and
  Werheid}{2014}]{folster2014facial}
F{\"o}lster, M.; Hess, U.; and Werheid, K.
\newblock 2014.
\newblock Facial age affects emotional expression decoding.
\newblock {\em Frontiers in Psychology}.

\bibitem[\protect\citeauthoryear{Goffman}{1959}]{goffman1959presentation}
Goffman, E.
\newblock 1959.
\newblock The presentation of self in everyday life.

\bibitem[\protect\citeauthoryear{Greenwood, Perrin, and
  Duggan}{2016}]{duggan2016social}
Greenwood, S.; Perrin, A.; and Duggan, M.
\newblock 2016.
\newblock Social media update 2016.
\newblock {\em Pew Research Center}.

\bibitem[\protect\citeauthoryear{Gross \bgroup et al\mbox.\egroup
  }{1997}]{gross1997emotion}
Gross, J.~J.; Carstensen, L.~L.; Pasupathi, M.; Tsai, J.; G{\"o}testam~Skorpen,
  C.; and Hsu, A.~Y.
\newblock 1997.
\newblock Emotion and aging: experience, expression, and control.
\newblock {\em Psychology and aging} 12(4):590.

\bibitem[\protect\citeauthoryear{Hogg}{2006}]{hogg2006social}
Hogg, M.~A.
\newblock 2006.
\newblock Social identity theory.
\newblock {\em Contemporary social psychological theories} 13:111--1369.

\bibitem[\protect\citeauthoryear{{International Markets Bureau,
  Canada}}{2012}]{mir2012}
{International Markets Bureau, Canada}.
\newblock 2012.
\newblock Global consumer trends - age demographics.
\newblock Available at \url{http://publications.gc.ca/pub?id=9.574883&sl=0},
  last accessed 8 March 2017.

\bibitem[\protect\citeauthoryear{Jain and Learned-Miller}{2010}]{jain2010fddb}
Jain, V., and Learned-Miller, E.~G.
\newblock 2010.
\newblock Fddb: A benchmark for face detection in unconstrained settings.
\newblock {\em UMass Amherst Technical Report}.

\bibitem[\protect\citeauthoryear{Jang \bgroup et al\mbox.\egroup
  }{2015}]{jang2015generation}
Jang, J.~Y.; Han, K.; Shih, P.~C.; and Lee, D.
\newblock 2015.
\newblock Generation like: Comparative characteristics in instagram.
\newblock In {\em CHI},  4039--4042.
\newblock ACM.

\bibitem[\protect\citeauthoryear{Jenkins and Burton}{2008}]{jenkins2008100}
Jenkins, R., and Burton, A.
\newblock 2008.
\newblock 100\% accuracy in automatic face recognition.
\newblock {\em Science} 319(5862):435--435.

\bibitem[\protect\citeauthoryear{Jia \bgroup et al\mbox.\egroup
  }{2014}]{jia2014caffe}
Jia, Y.; Shelhamer, E.; Donahue, J.; Karayev, S.; Long, J.; Girshick, R.;
  Guadarrama, S.; and Darrell, T.
\newblock 2014.
\newblock Caffe: Convolutional architecture for fast feature embedding.
\newblock {\em arXiv preprint arXiv:1408.5093}.

\bibitem[\protect\citeauthoryear{Krizhevsky, Sutskever, and
  Hinton}{2012}]{krizhevsky_imagenet_2012}
Krizhevsky, A.; Sutskever, I.; and Hinton, G.~E.
\newblock 2012.
\newblock Imagenet classification with deep convolutional neural networks.
\newblock In {\em NIPS},  1097--1105.

\bibitem[\protect\citeauthoryear{Lim \bgroup et al\mbox.\egroup
  }{2015}]{lim2015mytweet}
Lim, B.~H.; Lu, D.; Chen, T.; and Kan, M.-Y.
\newblock 2015.
\newblock \# mytweet via instagram: Exploring user behaviour across multiple
  social networks.
\newblock {\em ASONAM}.

\bibitem[\protect\citeauthoryear{Liu}{2007}]{liu2007social}
Liu, H.
\newblock 2007.
\newblock Social network profiles as taste performances.
\newblock {\em Journal of Computer-Mediated Communication} 13(1):252--275.

\bibitem[\protect\citeauthoryear{McCall and
  Simmons}{1978}]{mccall1978identities}
McCall, G., and Simmons, J.
\newblock 1978.
\newblock {\em Identities and interactions: An examination of associations in
  everyday life (revised ed.)}.
\newblock Free Press.

\bibitem[\protect\citeauthoryear{McClure}{2000}]{mcclure2000meta}
McClure, E.~B.
\newblock 2000.
\newblock A meta-analytic review of sex differences in facial expression
  processing and their development in infants, children, and adolescents.
\newblock {\em Psychological bulletin} 126(3):424.

\bibitem[\protect\citeauthoryear{McLaughlin and
  Vitak}{2012}]{mclaughlin2012norm}
McLaughlin, C., and Vitak, J.
\newblock 2012.
\newblock Norm evolution and violation on facebook.
\newblock {\em New Media \& Society} 14(2):299--315.

\bibitem[\protect\citeauthoryear{Mikolov \bgroup et al\mbox.\egroup
  }{2013a}]{mikolov2013efficient}
Mikolov, T.; Chen, K.; Corrado, G.; and Dean, J.
\newblock 2013a.
\newblock Efficient estimation of word representations in vector space.
\newblock In {\em Proceedings of Workshop at ICLR}.

\bibitem[\protect\citeauthoryear{Mikolov \bgroup et al\mbox.\egroup
  }{2013b}]{mikolov2013distributed}
Mikolov, T.; Sutskever, I.; Chen, K.; Corrado, G.~S.; and Dean, J.
\newblock 2013b.
\newblock Distributed representations of words and phrases and their
  compositionality.
\newblock In {\em Advances in neural information processing systems},
  3111--3119.

\bibitem[\protect\citeauthoryear{Mod}{2010}]{mod2010reading}
Mod, G. B.~B.
\newblock 2010.
\newblock Reading romance: The impact facebook rituals can have on a romantic
  relationship.
\newblock {\em Journal of comparative research in anthropology and sociology}
  (2):61--77.

\bibitem[\protect\citeauthoryear{{National Eye Institute,
  U.S.}}{2010}]{nei-stats}
{National Eye Institute, U.S.}
\newblock 2010.
\newblock Statistics and data: Myopia.
\newblock Available at \url{https://nei.nih.gov/eyedata/myopia}, last accessed
  8 March 2017.

\bibitem[\protect\citeauthoryear{National Eye~Institute}{2010}]{nei-facts}
National Eye~Institute, U.
\newblock 2010.
\newblock Facts about myopia.
\newblock Available at \url{https://nei.nih.gov/health/errors/myopia}, last
  accessed 8 March 2017.

\bibitem[\protect\citeauthoryear{Ozenc and Farnham}{2011}]{ozenc2011life}
Ozenc, F.~K., and Farnham, S.~D.
\newblock 2011.
\newblock Life modes in social media.
\newblock In {\em CHI},  561--570.
\newblock ACM.

\bibitem[\protect\citeauthoryear{Roberts}{2005}]{roberts2005changing}
Roberts, L.~M.
\newblock 2005.
\newblock Changing faces: Professional image construction in diverse
  organizational settings.
\newblock {\em Academy of management review} 30(4):685--711.

\bibitem[\protect\citeauthoryear{Rui and Stefanone}{2013}]{rui2013strategic}
Rui, J., and Stefanone, M.~A.
\newblock 2013.
\newblock Strategic self-presentation online: A cross-cultural study.
\newblock {\em Computers in Human Behavior} 29(1):110--118.

\bibitem[\protect\citeauthoryear{Sebastian and
  Bristow}{2008}]{sebastian2008formal}
Sebastian, R.~J., and Bristow, D.
\newblock 2008.
\newblock Formal or informal? the impact of style of dress and forms of address
  on business students' perceptions of professors.
\newblock {\em Journal of education for business} 83(4):196--201.

\bibitem[\protect\citeauthoryear{Sharda}{2009}]{sharda2009tourism}
Sharda, N.
\newblock 2009.
\newblock {\em Tourism Informatics: Visual Travel Recommender Systems, Social
  Communities, and User Interface Design: Visual Travel Recommender Systems,
  Social Communities, and User Interface Design}.
\newblock IGI Global.

\bibitem[\protect\citeauthoryear{Skeels and Grudin}{2009}]{skeels2009social}
Skeels, M.~M., and Grudin, J.
\newblock 2009.
\newblock When social networks cross boundaries: a case study of workplace use
  of facebook and linkedin.
\newblock In {\em GROUP},  95--104.
\newblock ACM.

\bibitem[\protect\citeauthoryear{Souza \bgroup et al\mbox.\egroup
  }{2015}]{souza15dawn}
Souza, F.; de~Las~Casas, D.; Flores, V.; Youn, S.; Cha, M.; Quercia, D.; and
  Almeida, V.
\newblock 2015.
\newblock Dawn of the selfie era: The whos, wheres, and hows of selfies on
  instagram.
\newblock In {\em ACM COSN}.

\bibitem[\protect\citeauthoryear{Strano}{2008}]{strano2008user}
Strano, M.~M.
\newblock 2008.
\newblock User descriptions and interpretations of self-presentation through
  facebook profile images.
\newblock {\em Cyberpsychology: Journal of Psychosocial Research on Cyberspace}
  2(2):5.

\bibitem[\protect\citeauthoryear{Tifferet and
  Vilnai-Yavetz}{2014}]{tifferet2014gender}
Tifferet, S., and Vilnai-Yavetz, I.
\newblock 2014.
\newblock Gender differences in facebook self-presentation: An international
  randomized study.
\newblock {\em Computers in Human Behavior} 35:388--399.

\bibitem[\protect\citeauthoryear{Tufekci}{2014}]{Tufekci14}
Tufekci, Z.
\newblock 2014.
\newblock Big questions for social media big data: Representativeness, validity
  and other methodological pitfalls.
\newblock In {\em AAAI {ICWSM}}.

\bibitem[\protect\citeauthoryear{Van~Dijck}{2013}]{van2013you}
Van~Dijck, J.
\newblock 2013.
\newblock `you have one identity': performing the self on facebook and
  linkedin.
\newblock {\em Media, Culture \& Society} 35(2):199--215.

\bibitem[\protect\citeauthoryear{Viola and Jones}{2004}]{viola2004robust}
Viola, P., and Jones, M.~J.
\newblock 2004.
\newblock Robust real-time face detection.
\newblock {\em Int'l. J. Comput. Vision} 57(2):137--154.

\bibitem[\protect\citeauthoryear{Walther \bgroup et al\mbox.\egroup
  }{2008}]{walther2008role}
Walther, J.~B.; Van Der~Heide, B.; Kim, S.-Y.; Westerman, D.; and Tong, S.~T.
\newblock 2008.
\newblock The role of friends’ appearance and behavior on evaluations of
  individuals on facebook: Are we known by the company we keep?
\newblock {\em Human Communication Research} 34(1):28--49.

\bibitem[\protect\citeauthoryear{Wright \bgroup et al\mbox.\egroup
  }{2009}]{wright2009robust}
Wright, J.; Yang, A.; Ganesh, A.; Sastry, S.; and Ma, Y.
\newblock 2009.
\newblock Robust face recognition via sparse representation.
\newblock {\em IEEE TPAMI} 31(2):210--227.

\bibitem[\protect\citeauthoryear{Zhao, Grasmuck, and
  Martin}{2008}]{zhao2008identity}
Zhao, S.; Grasmuck, S.; and Martin, J.
\newblock 2008.
\newblock Identity construction on facebook: Digital empowerment in anchored
  relationships.
\newblock {\em Computers in human behavior} 24(5):1816--1836.

\bibitem[\protect\citeauthoryear{Zhao, Lampe, and
  Ellison}{2016}]{zhao2016social}
Zhao, X.; Lampe, C.; and Ellison, N.~B.
\newblock 2016.
\newblock The social media ecology: User perceptions, strategies and
  challenges.
\newblock In {\em CHI},  89--100.
\newblock New York, NY, USA: ACM.

\bibitem[\protect\citeauthoryear{Zhong \bgroup et al\mbox.\egroup
  }{2014}]{zhong2014social}
Zhong, C.; Salehi, M.; Shah, S.; Cobzarenco, M.; Sastry, N.; and Cha, M.
\newblock 2014.
\newblock Social bootstrapping: how pinterest and last. fm social communities
  benefit by borrowing links from facebook.
\newblock In {\em WWW},  305--314.

\bibitem[\protect\citeauthoryear{Zhou \bgroup et al\mbox.\egroup
  }{2013}]{zhou2013extensive}
Zhou, E.; Fan, H.; Cao, Z.; Jiang, Y.; and Yin, Q.
\newblock 2013.
\newblock Extensive facial landmark localization with coarse-to-fine
  convolutional network cascade.
\newblock In {\em IEEE ICCV Workshops},  386--391.

\end{thebibliography}
}
\end{document}